# On the Elastic Stability of Folded Rings in Circular and Straight States

Sophie Leanza[a], Ruike Renee Zhao[a], and John W. Hutchinson[b]

[a]Dept. of Mechanical Engineering, Stanford University, Stanford, CA, 94305
[b]School of Engineering and Applied Sciences, Harvard University, Cambridge, MA, 02318

**Abstract:** Single-loop elastic rings can be folded into multi-loop equilibrium configurations. In this paper, the stability of several such multi-loop states which are either circular or straight are investigated analytically and illustrated by experimental demonstrations. The analysis ascertains stability by exploring variations of the elastic energy of the rings for admissible deformations in the vicinity of the equilibrium state. The approach employed is the conventional stability analysis for elastic conservative systems which differs from most of the analyses that have been published on this class of problems, as will be illustrated by reproducing and elaborating on several problems in the literature. In addition to providing solutions to two basic problems, the paper analyes and demonstrates the stability of six-sided rings that fold into straight configurations.[1]

## 1. Introduction

Slender elastic rods formed into rings can assume a surprising variety of equilibrium shapes other than the basic single-looped configuration. For example, a single-looped circular ring formed from a uniform straight rod can be twisted to form a ring having three equal sized circular loops with essentially no twist in the circular 3-loop state. Moreover, for rod cross-sectional geometries that have an out-of-plane bending stiffness large enough compared to the in-plane bending stiffness, both the 1-loop state and 3-loop state can be stable. The transition between the 1- and 3- loop configurations is three-dimensional, involving twist and out-of-plane displacements. This phenomenon has been known and exploited for many years, but other phenomena of this type are currently being uncovered and explored as illustrated by the recent articles by Wu et al. (2021, 2022), Lu et al. (2023), and Yu et al. (2023). The three-dimensional behavior of slender rod systems is of interest to a wide audience of scientists and mathematicians dealing with applications in mechanical and biological arenas. The motivation of the authors of the present paper derives primarily from structural applications of lightweight collapsible, or foldable, structures (Mouthuy et al. (2012); Leanza et al. (2022); Sun et al. (2022); Lachenal et al. (2012); Mhatre et al. (2021)).

---

[1] This paper is dedicated to Viggo Tvergaard, a major contributor to the subject of the stability of solids and structures, in celebration of his 80[th] birthday.

In the nineteenth century, Kirchhoff laid out the three-dimensional theory of elastic rods, generalizing Euler's two-dimensional elastica. Much of the work on this class of problems has been conducted within the framework Kirchhoff developed, including that presented in this paper. The main problem addressed in this paper is the stability of six-sided rings formed by joining six rods in the shape of 120° or 240° circular arcs into a complete ring. Each of the six arc segments has a uniform natural curvature $\kappa_n$ when unloaded. The rods are bent into 120° or 240° circular arcs and joined to form the ring as depicted in Fig. 1. Each rod segment joins its neighbor at a cusp. This 1-loop ring can be folded into a 3-loop ring with straight sides. The bending stresses in the folded state depend on the natural curvature and they are the source of the residual stresses in the structure. In this paper, the stability of the straight folded state is determined and, specifically, the largest natural curvature at which the straight state becomes unstable is obtained. Experimental demonstrations of the behavior of these rings will be presented.

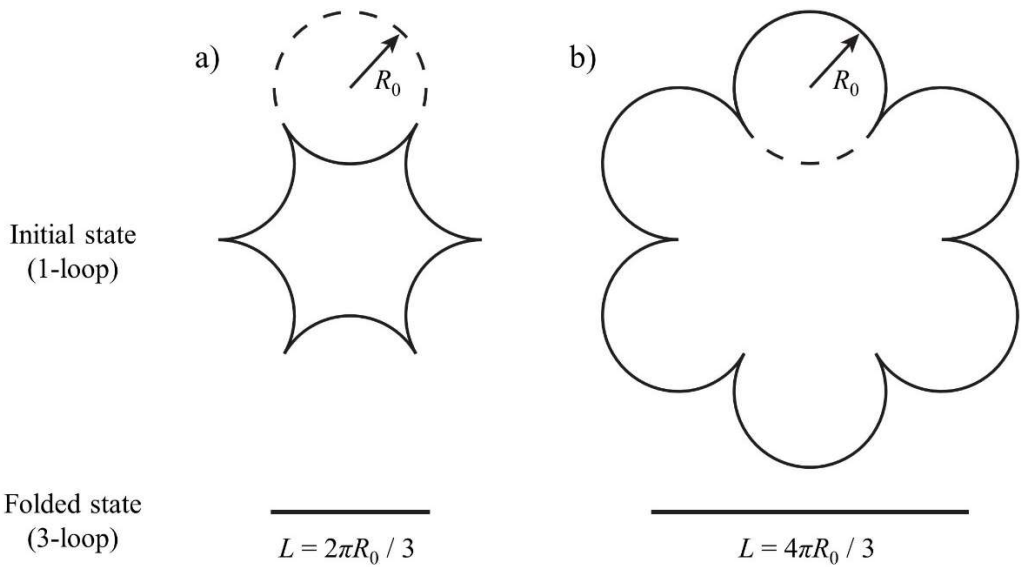

Fig. 1. a) A 6-sided ring formed from 120° circular arcs which folds into a straight 3-loop configuration of length $2\pi R_0 / 3$. b) A 6-sided ring formed from 240° circular arcs which folds into a 3-loop straight configuration of length $4\pi R_0 / 3$.

Many of the stability studies of three-dimensional rod structures, going back to some of the earliest papers, e.g., Michell (1889), are based on investigations of the vibration spectrum of the structure. These studies identify the load or residual stress level at which at least one natural frequency becomes zero. This approach is a natural choice because Kirchhoff rod theory is an Eulerian-type formulation with the variables describing the rod defined in the deformed state. For certain problems, such as those illustrated in this paper, an equivalent, but alternative, Lagrangian approach based on Kirchhoff theory may have advantages and may provide a more direct and less complicated pathway to ascertaining stability. In Section 2, equations are presented for the investigation of the stability of configurations that are either circular or straight. In Section 3, three problems previously analyzed in the literature will be re-analyzed to illustrate application of the approach adopted in this paper and to establish its validity. One of the three problems in Section 3 is the maximum curvature that can be imposed on a uniformly curved rod arc that is clamped at its ends. The importance of this problem will be emphasized, and detailed results not previously available in the literature will be presented. A model problem for a pre-stressed bi-rod is analyzed in Section 4 providing basic insights, and the stability of the 3-loop folded states of the six-sided rings are analyzed in Section 5.

## 2. Equations governing ring behavior for circular and straight reference configurations

The equations associated with deformations and strains based on Kirchhoff rod theory from circular reference states are first presented and then specialized to straight reference states. The procedures employed to derive these equations are the same as those given in many papers and books in the literature (Antman, 1995), and thus the present presentation will be as brief as possible. As seen in Fig. 2, the right hand set of unit vectors in the reference state, $(\mathbf{i}_r, \mathbf{i}_\theta, \mathbf{i}_3)$, have $\mathbf{i}_\theta$ tangent to the reference circle, $\mathbf{i}_r$ perpendicular to the circle, and $\mathbf{i}_3$ normal to the plane of the circle. The embedded set of unit vectors in the deformed state is denoted by $(\mathbf{e}_1, \mathbf{e}_2, \mathbf{e}_3)$ with $\mathbf{e}_2$ tangent to the rod and $(\mathbf{e}_1, \mathbf{e}_3)$ rotating with the cross-section. The set $(\mathbf{e}_1, \mathbf{e}_2, \mathbf{e}_3)$ coincides with $(\mathbf{i}_r, \mathbf{i}_\theta, \mathbf{i}_3)$ when the ring is in the reference state.

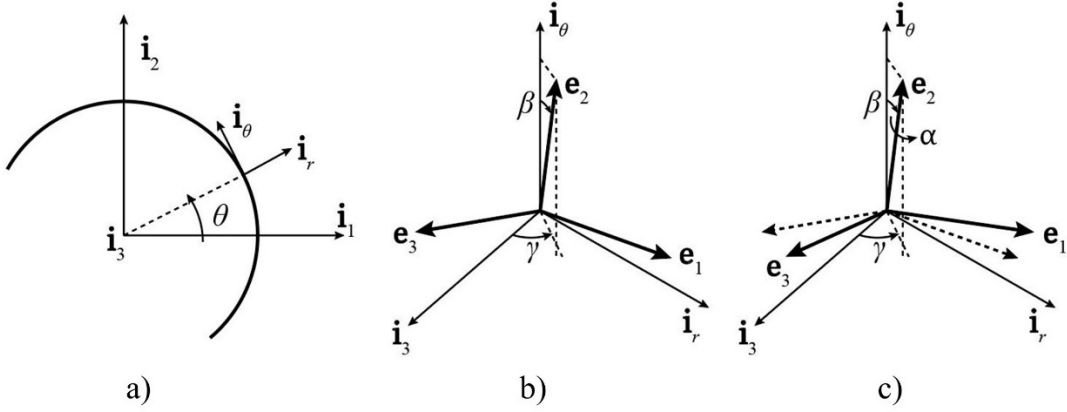

Fig. 2. Unit vectors and Euler angles. a) Cartesian base vectors and reference cylindrical base vectors. b) Local cylindrical base vectors and the two Euler angles, $\beta$ and $\gamma$, specifying the three embedded base vectors, $\mathbf{e}_i$, with no rotation about the rod axis, $\mathbf{e}_2$. c) Same as b) but after rotation $\alpha$ about $\mathbf{e}_2$ (dashed arrows denote $\mathbf{e}_1$ and $\mathbf{e}_3$ before rotation).

Euler angles are employed to describe the change of deformation of the rod for departures from the circular reference state. The two Euler angles that define the geometry of the axis of the rod are shown in Fig. 2b with no rotation of the embedded unit vectors about the axis. The third angle is the accumulated rotation of the cross-section about the $\mathbf{e}_2$-axis, $\alpha$, occurring during the motion from the circular state (Fig. 2c). Attention in this paper will be limited to inextensional rods, and the distance along the rod measured from some reference point and increasing with the circumferential angle $\theta$ is denoted by $s$. A Lagrangian description will be employed in the analyses in this paper. With $R$ as the radius of the ring, or rings, in the reference state, the unit vectors introduced above, $(\mathbf{i}_r(\theta), \mathbf{i}_\theta(\theta), \mathbf{i}_3)$, and the Euler angles, $(\alpha, \beta, \gamma)$, are functions of $\theta$ or $s$. The relations between the two sets of unit vectors are:

$$\begin{aligned}
\mathbf{e}_1 &= \left(\cos(\alpha+\gamma)+(1-\cos\beta)\sin\alpha\sin\gamma\right)\mathbf{i}_r + \sin\alpha\sin\beta\ \mathbf{i}_\theta \\
&\quad + \left(-\sin(\alpha+\gamma)+(1-\cos\beta)\sin\alpha\cos\gamma\right)\mathbf{i}_3 \\
\mathbf{e}_2 &= \sin\beta\sin\gamma\,\mathbf{i}_r + \cos\beta\,\mathbf{i}_\theta + \sin\beta\cos\gamma\,\mathbf{i}_3 \\
\mathbf{e}_3 &= \left(\sin(\alpha+\gamma)-(1-\cos\beta)\cos\alpha\sin\gamma\right)\mathbf{i}_r - \cos\alpha\sin\beta\,\mathbf{i}_\theta \\
&\quad + \left(\cos(\alpha+\gamma)-(1-\cos\beta)\cos\alpha\cos\gamma\right)\mathbf{i}_3
\end{aligned} \qquad (2.1)$$

The direction cosines between the axes, $l_{ij} = \mathbf{i}_i \bullet \mathbf{e}_j$ such that $\mathbf{e}_i = l_{ji}\mathbf{i}_j$, with $(\mathbf{i}_r, \mathbf{i}_\theta) \to (\mathbf{i}_1, \mathbf{i}_2)$ for notational convenience, and $\mathbf{i}_i = l_{ij}\mathbf{e}_j$. These are provided in the Supplementary Materials.

Denote the displacement components from the circular reference state by $\mathbf{u} = u_r\mathbf{i}_r + u_\theta\mathbf{i}_\theta + u_3\mathbf{i}_3$. A vector tangent to the centerline axis pointing in the direction of $\mathbf{e}_2$ is

$$\mathbf{t} = (du_r/ds - R^{-1}u_\theta)\mathbf{i}_r + (1 + du_\theta/ds + R^{-1}u_r)\mathbf{i}_\theta + du_3/ds\,\mathbf{i}_3 \equiv -\varphi_3\mathbf{i}_r + (1+\varepsilon_\theta)\mathbf{i}_\theta + \varphi_r\mathbf{i}_3$$

The inextensionality condition is

$$2(du_\theta/ds + R^{-1}u_r) + (du_\theta/ds + R^{-1}u_r)^2 + (du_r/ds - R^{-1}u_\theta)^2 + (du_3/ds)^2 = 0 \qquad (2.2)$$

which also ensures that $\mathbf{t} = \mathbf{e}_2$ is a unit vector. Note that $\varepsilon_\theta \equiv (du_\theta/ds + R^{-1}u_r)$, $\varphi_r = du_3/ds$ and $\varphi_3 = -(du_r/ds - R^{-1}u_\theta)$ are, respectively, the centerline linearized stretching strain and linearized rotations about $\mathbf{i}_r$ and $\mathbf{i}_3$. Furthermore, solving the inextensionality condition for $\varepsilon_\theta$ in terms of the other two quantities (with $\varphi_r^2 + \varphi_3^2 < 1$) gives the constraint expressed as

$$\varepsilon_\theta = \sqrt{1 - \varphi_r^2 - \varphi_3^2} - 1 \qquad (2.3)$$

By comparing the terms in the equation for $\mathbf{t}$ with those for $\mathbf{e}_2$, the following exact relations are also seen:

$$\varepsilon_\theta = (du_\theta/ds + R^{-1}u_r) = \cos\beta - 1 \qquad (2.4)$$

or $\cos\beta = \sqrt{1 - \varphi_r^2 - \varphi_3^2}$ and

$$\varphi_r = du_3/ds = \sin\beta\cos\gamma \text{ and } \varphi_3 = -(du_r/ds - R^{-1}u_\theta) = -\sin\beta\sin\gamma \qquad (2.5)$$

If $\beta \neq 0$, it also follows that

$$\tan\gamma = -\varphi_3/\varphi_r \qquad (2.6)$$

With $\kappa_1$ and $\kappa_3$ as the curvatures about the $\mathbf{e}_1$ and $\mathbf{e}_3$ axes, respectively, and $\kappa_2$ as the derivative of twist with respect to distance along the rod, one can use the Frenet-Seret formula (c.f., Champneys and Thompson, 1996; Audoly and Seffen, 2015),

$$\kappa_i = \frac{1}{2}\varepsilon_{ijk}\frac{d\mathbf{e}_j}{ds}\bullet\mathbf{e}_k, \qquad (2.7)$$

to compute the curvatures. The exact results are

$$\begin{aligned}
\kappa_1 &= \cos\alpha\,(d\beta/ds - R^{-1}\sin\gamma) - \sin\alpha(-\sin\beta\,d\gamma/ds + R^{-1}\cos\beta\cos\gamma) \\
\kappa_2 &= d\alpha/ds + \cos\beta\,d\gamma/ds + R^{-1}\sin\beta\cos\gamma \\
\kappa_3 &= \cos\alpha\,(-\sin\beta\,d\gamma/ds + R^{-1}\cos\beta\cos\gamma) + \sin\alpha(d\beta/ds - R^{-1}\sin\gamma)
\end{aligned} \quad (2.8)$$

The curvature components are defined with respect to the embedded unit vector system $\mathbf{e}_i$ in the deformed state, as reflected by the vector, $\mathbf{\Omega} = \kappa_i \mathbf{e}_i$.

For the stability bifurcation calculations to follow, the 'natural' uncoiled state of the rod has length $L$ and an initial uniform curvature $\kappa_n$ about the 3-axis. Attention here is restricted to uniform rods with linear bending and twisting properties. The principal axes of the cross-section in the natural state are taken to be aligned with $(\mathbf{e}_1, \mathbf{e}_3)$, i.e., $(\mathbf{i}_r, \mathbf{i}_3)$ in the reference state, and the bending stiffnesses about these two axes are denoted by $B_1$ and $B_3$. The torsional stiffness about the $\mathbf{e}_2$-axis is denoted by $B_2$. For all but one of the problems analyzed in this paper, the rods in their unstressed state have a uniform natural curvature, $\kappa_n$, about the $\mathbf{i}_3$-axis. Then, the rods are bent uniformly about the 3-axis into circular loops or arc segments of radius $R$. Stability or lack thereof is investigated in this circular state of radius. We will also investigate stability of straight states with $R \to \infty$. In the circular reference state, the non-zero curvature is $\kappa_3 = R^{-1}$, the change of curvature from the natural state is $R^{-1} - \kappa_n$, and the moment in the ring in the circular state is $M_3 = B_3(R^{-1} - \kappa_n)$. For rods of length $L$ in reference states that are either circular or straight, the strain energy in configurations displaced from the reference state is assumed to be governed by linear constitutive behavior such that

$$SE = \tfrac{1}{2}\int_0^L \left\{ B_1\kappa_1^2 + B_2\kappa_2^2 + B_3(\kappa_3 - \kappa_n)^2 \right\} ds \quad (2.9)$$

with curvatures given by (2.8). These curvature expressions hold for reference states that are straight by setting the terms multiplied by $R^{-1}$ to zero. Equations (2.8) agree with those given by Champneys and Thompson (1996) in their study of initially straight rods, but with different notation.

*2.1 Moderate rotation approximation*

For moderate rotations from the circular state (with no initial twist) we anticipate that $\alpha^2$, $\varphi_r^2$ and $\varphi_3^2$ are all small compared to 1. Bifurcation calculations of stability of the circular or straight state require the energy changes of the elastic system to be accurate to the second order in the displacements and their gradients from that state. It follows from the equations given above that $\beta^2$ is also small compared to 1, however, in the circular state, $\gamma$ cannot be inferred from $\tan\gamma = -\varphi_3/\varphi_r$. In the problems investigated here, the curvatures must vary continuously from those in the reference state as bifurcation proceeds from this state. When there is no initial torsion in reference state, the formulas for $\kappa_1$ and $\kappa_3$ in the circular state with $(\alpha = 0, \beta = 0)$ reveal that $\gamma$ must be zero at bifurcation, and it must emerge continuously from zero given the initial curvatures $(\kappa_1 = 0, \kappa_3 = R^{-1})$ and a continuous variation of the curvatures. Assuming $(\alpha, \beta, \gamma)$ are all small and expanding the curvature expressions in these quantities in a Taylor series up to and including quadratic terms one obtains:

$$\kappa_1 = \frac{d\beta}{ds} - R^{-1}\omega + O_3, \quad \kappa_2 = \frac{d\omega}{ds} + R^{-1}\beta + O_3,$$

$$\kappa_3 = R^{-1}\left(1 - \frac{1}{2}\omega^2 - \frac{1}{2}\beta^2\right) - \beta\frac{d\omega}{ds} + \frac{d(\alpha\beta)}{ds} + O_4 \quad (2.10)$$

with $\omega = \alpha + \gamma$. The notation $O_3$ and $O_4$ signifies that no terms below cubic and quartic order, respectively, have been neglected. Thus, the first two curvatures have no quadratic terms, and the third curvature has no cubic terms. To the same order,

$$\varepsilon_\theta = \left(\frac{du_\theta}{ds} + R^{-1}u_r\right) = -\frac{1}{2}\beta^2 = -\frac{1}{2}(\varphi_r^2 + \varphi_3^2), \quad \varphi_r = \frac{du_3}{ds} = \beta, \quad \varphi_3 = \left(\frac{du_r}{ds} - R^{-1}u_\theta\right) = -\beta\gamma \quad (2.11)$$

*2.2 The strain energy functional to quadratic order*

Up to and including quadratic terms, the strain energy (2.9) is

$$SE = \tfrac{1}{2}B_3(1/R - \kappa_n)^2 L + P_2(\beta, \omega, M) \quad (2.12)$$

where

$$P_2 = \tfrac{1}{2}\int_0^L \left\{ B_1\left(\frac{d\beta}{ds}-\frac{\omega}{R}\right)^2 + B_2\left(\frac{d\omega}{ds}+\frac{\beta}{R}\right)^2 - B_3 M\left(\left(\frac{\beta}{R}\right)^2+\left(\frac{\omega}{R}\right)^2+2\frac{\beta}{R}\frac{d\omega}{ds}\right)\right\}ds + 2B_3 M\left.\frac{\alpha\beta}{R}\right|_0^L$$

with $\omega = \alpha + \gamma$, and $M = 1 - R\kappa_n$, the parameter introduced by Audoly and Seffen (2015). Because of the symmetry of the ring or arc, and the accuracy of the curvature expressions noted earlier, $SE$ has no cubic terms, so, in fact, (2.12) is accurate to $O_4$.

The quadratic functional $P_2$ determines the second variation of $SE$ about the reference state and governs its stability. Consider either complete rings, for which $\alpha$, $\gamma$ and $\beta$ are continuous across $s = 0$ and $s = L$, or circular arcs clamped at both ends ($\alpha = \beta = 0$). For these entities, $\alpha\beta|_0^L = 0$. The functional $P_2$ defines an eigenvalue problem with $M = 1 - R\kappa_n$ as the eigenvalue. To the order relevant to the bifurcation problem, the inextensionality condition requires $\varepsilon_\theta = (du_\theta / ds + u_r / R) = 0$. Additional conditions imposed on the Euler angles by displacement constraints may have to be taken into consideration, as will be illustrated. If $M = 0$, $P_2 \geq 0$ for all nonzero variations $(\beta, \omega)$. $P_2$ only vanishes for nonzero variations having $\omega = \beta'$ and $\omega' + \beta = 0$ which correspond to linearized rigid-body rotations. In other words, if $M = 0$, $P_2 > 0$ for all non-zero admissible variations that are not rigid-body motions. For complete rings, the Euler angles and their associated displacements in the bifurcation problem are required to be continuous and differentiable with period $L = 2\pi NR$ where $N$ is the number of loops, and continuity of displacement around the ring requires $\int_0^L \beta ds = 0$. For either complete rings or arcs, we seek the range of $M$, such that $P_2 > 0$ for all nonzero admissible variation (excluding rigid-body motions. The circular state is stable within this range and unstable for $M$ outside this range. For complete rings, the class of deformations involving a uniform axisymmetric twist, i.e., $\alpha = \omega$ independent of $\theta$ with $\beta = \gamma = 0$, deserves special attention. The integrand of $P_2$ for this class of deformations is simply $(b_1 - M)\alpha^2$, and thus the ring is unstable against uniform axisymmetric twist if $M > b_1$ and stable if $M < b_1$. Further discussion of this mode will be given later.

**3. Three solved problems in the literature**

In this section we use the standard quasi-static approach to the stability, or buckling, of elastic conservative systems outlined above to reproduce solutions to three problems previously analyzed and solved in the literature. The Lagrangian quasi-static method employed here, in some cases, generates results more directly. In the case of the second problem, the stability of pure bending of clamped circular and straight arcs, we will provide more complete and extensive results than are available in the literature and follow up in Sections 4 and 5 by demonstrating the importance of this fundamental problem.

*3.1 The stability of circular rings untwisted in 1-loop or 3-loop states*

The remarkable property of a 1-loop elastic circular ring to deform into multi-loop circular rings without twist when the number of loops is odd ($N = 3, 5,...$) has been known for years (Goto et al., 1992, Stojanoska and Stoytchev, 2008), but this phenomena is evidently not widely known to the community concerned with the buckling of structures. Two papers have provided a rather complete analysis of multi-state rings within the framework of Kirchhoff rod theory dealing with both the stability of the circular states and the transitional behavior linking them (Manning and Hoffman, 2001, Audoly and Seffen, 2015). The present paper will analyze the stability of circular and straight states but not the transition between the states. Our results reproduce some those in the studies of Audoly and Seffen and those still earlier by Manning and Hoffman. This section serves the purpose of presenting and validating the present approach for a solved problem.

We seek the critical eigenvalues, $M_C^-$ and $M_C^+$, and the associated modes, that define the range of stability of the ring subject to the condition that $\beta(\theta)$ and $\omega(\theta)$, with $\theta = s/R$, have periodicity $2\pi N$ with continuous derivatives. The eigenvalue equations are obtained from rendering the variations of $P_2$ stationary with respect to the variables $(\beta, \omega)$. The ordinary differential equations (ODEs) governing the eigenvalue problem are

$$b_2\omega'' + (M - b_1)\omega + (-M + b_1 + b_2)\beta' = 0 \\ b_1\beta'' + (M - b_2)\beta - (-M + b_1 + b_2)\omega' = 0 \tag{3.1}$$

with $b_1 = B_1/B_3$, $b_2 = B_2/B_3$, and $(\ )' = d(\ )/d\theta$. These equations do not involve $\varphi_3$ or the in-plane displacements, $u_r$ and $u_\theta$. From (2.11), these quantities are of second order and play no role in the first order bifurcation problem, which involves a combination of the out-of-plane displacement, $u_3' = R\beta$, and twist. Continuity of $u_3$ requires $\int_0^{2N\pi} \beta d\theta = 0$.

A central concern of Audoly and Seffen (2015) is ribbon-like rods with high aspect ratio cross-sections for which the bending-torsional behavior is nonlinear due to developable deformations of the ribbon. Throughout this paper the emphasis will be on rods with linear bending-torsional constitutive behavior. The height to thickness ratios of the cross sections are assumed to be in the range such that ribbon-like nonlinear behavior does not occur. Consider modes of the form

$$\omega = A\cos(j\theta/N), \quad \beta = B\sin(j\theta/N) \tag{3.2}$$

where $j$ is any non-negative integer. The modal displacement is $u_3 = -B(RN/j)\cos(j\theta/N)$ for $j > 0$ and these all satisfy $\int_0^{2N\pi} \beta d\theta = 0$. (Potential modes of the form $\beta = const.$ are periodic but they are excluded by the condition $\int_0^{2N\pi} \beta d\theta = 0$.) The algebraic eigenvalue problem is

$$A\left[b_1 + b_2(j/N)^2 - M\right] - B\left[(b_1 + b_2 - M)(j/N)\right] = 0$$
$$-A\left[(b_1 + b_2 - M)(j/N)\right] + B\left[b_1(j/N)^2 + b_2 - M\right] = 0$$

leading to

$$\left(M^2 - M(b_1 + b_2) + b_1 b_2 \left(1 - (j/N)^2\right)\right)\left(1 - (j/N)^2\right) = 0$$

whose solutions are $j/N = 1$ for any $M$ (rigid-body modes), and

$$(M^+, M^-) = \frac{1}{2}(b_1 + b_2) \pm \sqrt{\frac{1}{4}(b_1 - b_2)^2 + b_2(j/N)^2}, \tag{3.3}$$

for all other $j$ and $N$. This agrees with the solution of Audoly and Seffen (2015) and Manning and Hoffman (2001)

For demonstrating and validating the present method, in this subsection we will consider only rings with an *isotropic cross-section* made of an isotropic elastic material. For circular cross-sections, solid or annular, these have $b_1 = 1$ and $b_2 = 1/(1+\nu)$ with $\nu$ as Poisson's ratio. The uniform axisymmetric twisting mode for any integer number of loops, corresponding to $j = 0$, with $\omega = A$, $\beta = 0$, $M^+ = 1$ and $M^- = b_2$ has been discussed earlier. Rods with cross-sections having $b_1 = 1$ are unstable in the uniform twisting mode if $M > M^+ = 1 - R\lambda_n > 1$, i.e., $\lambda_n < 0$. If the rod in the natural state is straight, i.e., $\lambda_n = 0$, it is easy to see that strain energy of the ring does not change subject to uniform axisymmetric twist. The ring has neutral stability with respect to this class of deformations. Any such rod with an initial curvature $\lambda_n < 0$ would immediately undergo a 180° uniform axisymmetric twist when shaped into a ring, thus effectively switching the sign of the initial natural curvature. In other words, the uniform axisymmetric twisting mode rules out the necessity of further consideration of rings of isotropic bending stiffness formed from rods with a negative natural curvature. The shape of the ring is unchanged by uniform axisymmetric twist, but the material of which it is composed undergoes deformation changes. The second eigenvalue, $M^- = b_2$, is spurious because it is associated with the trivial eigenmode, $\beta = \omega = 0$.

The modes associated with $j/N = 1$ for any $M$ are rigid-body rotations about axes in the plane of the ring. For the form assumed, $B = A$ such that $\omega = A\cos\theta$, $\varphi_r = \beta = A\sin\theta$, $\varphi_3 = 0$, $u_3 = -AR\cos\theta$, and $u_r = u_\theta = 0$. The one exception is the additional possibility of a mode involving deformation of the ring if $M = M^+ = 1 + b_2$, but this will not be consequential because the associated natural curvature, $R\kappa_n = -b_2$, is negative and superseded by uniform axisymmetric twist.

For 1-loop rings ($N = 1$) and $j \geq 2$, the range of stability for the $j^{th}$ mode is $M^- < M < M^+$. All the values of $M^+$ given by (3.3) satisfy $M^+ > 1$ and thus the upper limit of the stability range is determined by uniform axisymmetric torsion and $M_C^+ = 1$. The largest value of $M^-$ is associated with $j = 2$ and (3.3) gives $M_C^- = -0.861$ for circular cross-sections with $\nu = 1/3$. Thus for 1-loop rings, the stability range is $-0.861 < M < 1$. Consider rings formed from a rod of length $L$ with natural curvature $\kappa_n$ with either one or multiple loops such that $R = L/(2\pi N)$. This is the relevant normalization for rings formed from a rod with a given length and natural curvature that assumes one or more loops. The stability limits of the natural curvature are given by

$$\frac{L}{2\pi}\kappa_n^- = N(1-M_C^+) \quad \text{and} \quad \frac{L}{2\pi}\kappa_n^+ = N(1-M_C^-)$$

For the 1-loop ring, the range of the natural curvature for which the circular state is stable is

$$0 < (L/2\pi)\kappa_n < 1.86 \quad (N=1)$$

Determination of stability of the circular state for the same ring with multiple loops requires an additional consideration. For each of the modes with $j/N > 1$, the stability range is again given by $M^- < M < M^+$, but for each of the modes with $0 < j/N < 1$, the stability range is $M < M^-$, with $M^+$ playing no role. This is easily established by showing that for $M < M^-$, $P_2 \geq 0$ for all $A$ and $B$, while negative values of $P_2$ exist for some combinations of $A$ and $B$ for every $M$ if $M > M^-$. For the 3-loop ring ($N=3$), the largest value of $M$ for stability is set by $j=2$ and given by $M^- = 0.284$. The largest lower stability limit of $M$ is determined by $M_C^- = -0.286$ for $j=4$. Thus, the stability range for $N=3$ is $-0.286 < M < 0.284$ such that

$$2.15 < (L/2\pi)\kappa_n < 3.86 \quad (N=3)$$

The stability ranges of the 1- and 3- loop rings do not overlap for circular cross-sections. If the natural state of the rod is straight, the 1-loop circular state is marginally stable (susceptible to uniform axisymmetric torsion), however the 3-loop state is not. While it is possible to deform a stable 1-loop ring into the 3-loop circular state, that state would not be stable. Conversely, if the natural curvature is such that the 3-loop ring is stable, it would not transition to a circular ring with one loop when dislodged from the 3-loop state. Rods with cross-section having sufficiently large out-of-plane bending stiffness compared to in-plane stiffness have overlapping stability ranges for 1- and 3-loop rings. The modal displacement, $u_3 = -B(RN/j)\cos(j\theta/N)$, with the associated twist (3.2), is a progressive separation of the loops around the ring which does not bring the ring into contact with itself.

*3.2 Stability of pure bending of straight or naturally curved rods clamped at their ends*

Consider another problem analyzed by Audoly and Seffen (2015): the stability of pure bending about the $\mathbf{i}_3$-axis of rods with natural curvature. We will see in Sections 4 and 5 that this basic problem is fundamental to more complex structural stability phenomena. Here, reasonably complete results for rods with circular or rectangular cross-sections will be presented. As depicted in Fig. 3, the natural state of the rod is a uniform curvature, $\kappa_n > 0$, and the deformed state can be either curved ($\kappa = 1/R$) or straight ($\kappa = 1/R = 0$). The rod is inextensional and of length $L$. Take $s = R\theta$ measured from the center of the rod such that its ends are at $s = \pm L/2$. The rod is subject to pure bending by equal and opposite end moments of magnitude $B_3 |\kappa - \kappa_n|$ with $\kappa = 1/R$ as the curvature in the deformed state. In this state, the ends of the beam are clamped against further rotation. The stability of this state requires $P_2 > 0$ in (2.12) for all non-zero admissible variations from the current circular state. Due to the clamped end conditions, $\alpha\beta\big|_0^L = 0$. To remove the constraint imposed by the requirement $\int_{-L/2}^{L/2} \beta ds = 0$, substitute $\beta = du_3/ds$ in $P_2$. With $(u_3, \omega)$ as unknowns, the boundary conditions at each clamped end of the rod are $u_3 = du_3/ds = \omega = 0$.

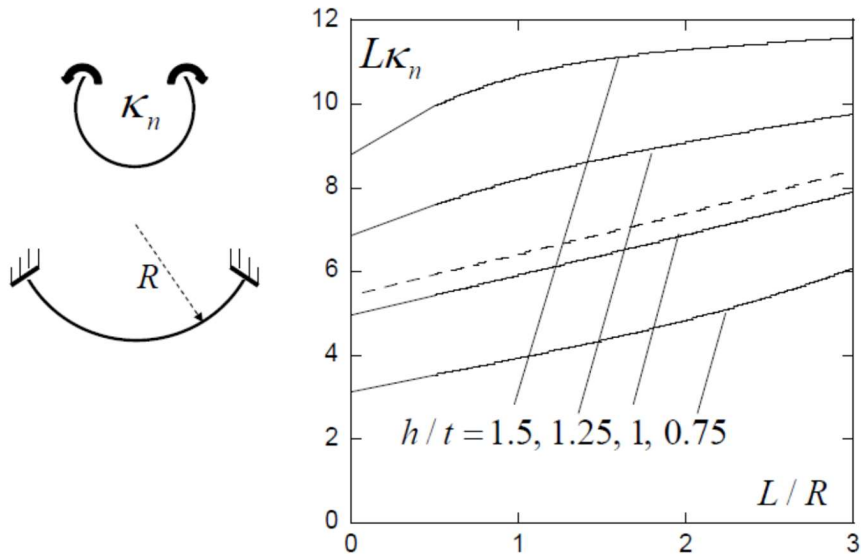

Fig. 3. Stability range of the natural curvature for a rod of length $L$ with either a circular (dashed line) or a rectangular (solid lines, depending on $h/t$) cross-section that is bent into an arc of radius $R$ and clamped at its ends. For natural curvatures greater than those plotted, the arc will buckle. Poisson's ratio is 1/3 in this plot. The limit in which the bent rod is straight, $L/R \to 0$, is given by (3.9). For the bent straight rod, the result applies when there is no constraint on the displacement parallel to the rod at the right end.

With $x = s/L$, $u = u_3/L$, $M = 1 - R\kappa_n$, and $\bar{L} = L/R$, the ODEs obtained by rendering $P_2$ in (2.12) stationary are

$$b_1(u'''' - \bar{L}\omega''') - b_2\bar{L}(\omega'' + \bar{L}u'') + M\bar{L}(\bar{L}u'' + \omega'') = 0$$
$$b_1\bar{L}(u'' - \bar{L}\omega) + b_2(\omega''' + \bar{L}u''') + M\bar{L}(\bar{L}\omega - u'') = 0 \quad (3.4)$$

The general solution to these equations can be split into symmetric ($c_1$, $c_2$, $c_3$) and anti-symmetric ($c_4$, $c_5$, $c_6$) parts as

$$u = c_1 + c_2 \cos(\bar{L}x) + c_3 \cos(\lambda\bar{L}x) + c_4 x + c_5 \sin(\bar{L}x) + c_6 \sin(\lambda\bar{L}x)$$
$$\omega = -\bar{L}\left(c_2 \cos(\bar{L}x) + c_3 q \cos(\lambda\bar{L}x) + c_5 \sin(\bar{L}x) + c_6 q \sin(\lambda\bar{L}x)\right) \quad (3.5)$$

with (anticipating $(b_1 - M)(b_2 - M) > 0$)

$$\lambda = \sqrt{\frac{(b_1 - M)(b_2 - M)}{b_1 b_2}}, \quad q = \frac{b_2 - M}{b_2} \quad (3.6)$$

The eigenvalue equations obtained by satisfying the boundary conditions at $x = 1/2$ are

$$\lambda \sin(\lambda\bar{L}/2)\cos(\bar{L}/2) - q\cos(\lambda\bar{L}/2)\sin(\bar{L}/2) = 0 \quad \text{(symmetric)} \quad (3.7)$$

$$2(1-q)\sin(\bar{L}/2)\sin(\lambda\bar{L}/2) - \lambda\bar{L}\sin(\bar{L}/2)\cos(\lambda\bar{L}/2) \\ + q\bar{L}\cos(\bar{L}/2)\sin(\lambda\bar{L}/2) = 0 \quad \text{(anti-symmetric)} \quad (3.8)$$

Before discussing some details of this solution, we first present the results for what will prove to be an important limiting case—the limit when the clamped rod is straight, $R \to \infty$ and $\bar{L} = L/R \to 0$. To the order relevant to the bifurcation analysis, $u_2 = 0$, and the combination of inextensionality and fully clamped conditions at both ends of the rod is still applicable in this limit. Obviously, a straight rod that is inextensional and clamped in this manner could not deflect from straight state. However, if, for example, the support condition at the right end of the rod is modified slightly so that the rod displacement is unconstrained in the 2-direction but otherwise constrained against rotation and displacement, the rod could undergo buckling displacements at the critical condition. The present analysis is equally applicable with this slight modification (similarly, if in the circular state one end of the rod is free to displace in the circumferential direction, the analysis in this section still applies). The results for the rod in the straight state can be obtained either by taking the limit of the above equations or by starting from scratch using the ODEs (3.4) with $\bar{L}=0$, noting that $M\bar{L} =\to -L\kappa_n$. One finds that the critical mode is symmetric and the maximum natural curvature for which the straight clamped rod is stable is

$$L\kappa_n = 2\pi \frac{\sqrt{B_1 B_2}}{B_3} \tag{3.9}$$

The associated mode is a combination of displacement and rotation given by

$$u_3 = cL\left(1+\cos(2\pi s/L)\right), \quad \omega = -c2\pi\sqrt{B_1/B_2}\left(1+\cos(2\pi s/L)\right) \tag{3.10}$$

The dimensionless combination of bending and twisting stiffnesses, $\sqrt{B_1 B_2}/B_3$ governing the critical natural curvature is plotted in Fig. 4 for rods with elliptical and rectangular cross-sections each having height $h$ in the 3-direction and thickness $t$ in the 1- or $r$-direction. Recall that for circular cross-sections (solid or annular), $b_1 = B_1/B_3 = 1$ and $b_2 = B_2/B_3 = 1/(1+\nu)$ so that $\sqrt{B_1 B_2}/B_3 = 1/\sqrt{1+\nu}$. For elliptical cross-sections

$$\frac{\sqrt{B_1 B_2}}{B_3} = \frac{h}{t}\sqrt{\frac{2}{(1+\nu)(1+(t/h)^2)}} \tag{3.11}$$

For solid rectangular cross-sections, $b_1 = (h/t)^2$, and an accurate approximation for $b_2$ is obtained using Sokolnikoff's (1956) result for the torsional stiffness:

$$b_2 = \frac{B_2}{B_3} = \frac{2}{1+\nu}\left(1 - \frac{195}{\pi^5}\frac{t}{h}\tanh\left(\frac{\pi}{2}\frac{h}{t}\right)\right) \tag{3.12}$$

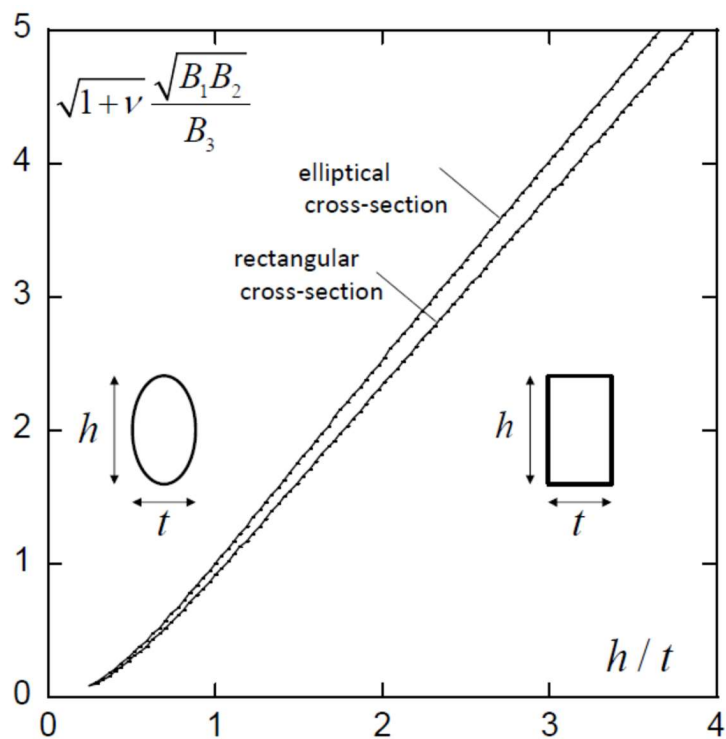

Fig. 4. A dimensionless combination of the bending and torsional stiffnesses important to rod stability (see 3.9) for rectangular and elliptical cross-sections.

The stable range of natural curvature for rods of circular and rectangular cross-sections is presented in Fig. 3. For all cases plotted, the anticipated condition, $(b_1 - M)(b_2 - M) > 0$, is satisfied and the symmetric mode with eigenvalue given by (3.7) is critical. Although the circular cross-section and the square cross-section both have $b_1 = 1$, $b_2$ is slightly smaller for the square cross-section and that is reflected in its slightly smaller range of stability. Not surprisingly, the aspect ratio of the rectangular cross-section, $h/t$, has a very strong influence on the stability range. One case for a cross-section with higher in-plane bending stiffness than out-of-plane stiffness, $h/t = 0.75$, is included in Fig. 3 showing that rods having such cross-sections have a dramatically reduced stability range. When the aspect ratio is very high, e.g., $h/t \gg 1$, (3.11) predicts a large range of stability. Such cross-sections are susceptible to the ribbon-like instabilities modeled by Audoly and Seffen (2015).

The eigenvalue conditions (3.7) and (3.8) have two branches. The one plotted in Fig. 3. and in the formula for a straight rod in (3.9) is for $\kappa_n > 1/R$. The other branch yields values of $\kappa_n < 1/R$. For example, for the straight rod, this second branch has $\kappa_n = -2\pi\sqrt{B_1 B_2}/B_3$, as obvious from symmetry.

Consider another fundamental problem governed by the eigenvalue equations (3.4): *The maximum length $L$ of a straight rod ($\kappa_n = 0$) that can be bent into a clamped arc of radius $R$ such that the rod does not buckle.* Equations (3.4) apply with $M = 1$ and clamped conditions $u = u' = \omega = 0$ at the two ends of the rod. We omit the details of this analysis, which is similar to that already given. The mode associated with the critical value of $\bar{L} = L/R$ at which the bent rod loses stability is found to be symmetric about center of the rod with eigenvalues given by:

$$\lambda \sin(\lambda \bar{L}/2)\cos(\bar{L}/2) - q\cos(\lambda \bar{L}/2)\sin(\bar{L}/2) = 0, \quad \text{if } (b_1-1)(b_2-1) > 0 \qquad (3.13)$$

with $\lambda = \sqrt{(b_1-1)(b_2-1)/b_1 b_2}$ and $q = (b_2-1)/b_2$, and

$$\lambda \sinh(\lambda \bar{L}/2)\cos(\bar{L}/2) + q\cosh(\lambda \bar{L}/2)\sin(\bar{L}/2) = 0, \quad \text{if } (b_1-1)(b_2-1) < 0 \qquad (3.14)$$

with $\lambda = \sqrt{-(b_1-1)(b_2-1)/b_1 b_2}$ and $q = (b_2-1)/b_2$. The largest normalized length, $L/R$, for which the clamped bent arc is stable is plotted in Fig. 5 as a function of $h/t$ for rods of rectangular cross-section. Any rod with equal bending stiffnesses, i.e., $b_1 = 1$, so that $\lambda = 0$ in (3.13) and (3.14), has a critical length given by $L/R = 2\pi$, with an associated mode that is a combination of uniform axisymmetric torsion and rigid body rotation such that there is no rotation where the ring is clamped. The critical stability limit is extremely sensitive to the cross-section aspect ratio for rods with in-plane and out-of-plane bending stiffnesses that are nearly the same. A slight increase of the out-of-plane bending stiffness relative to the in-plane stiffness significantly increases the stability limit.

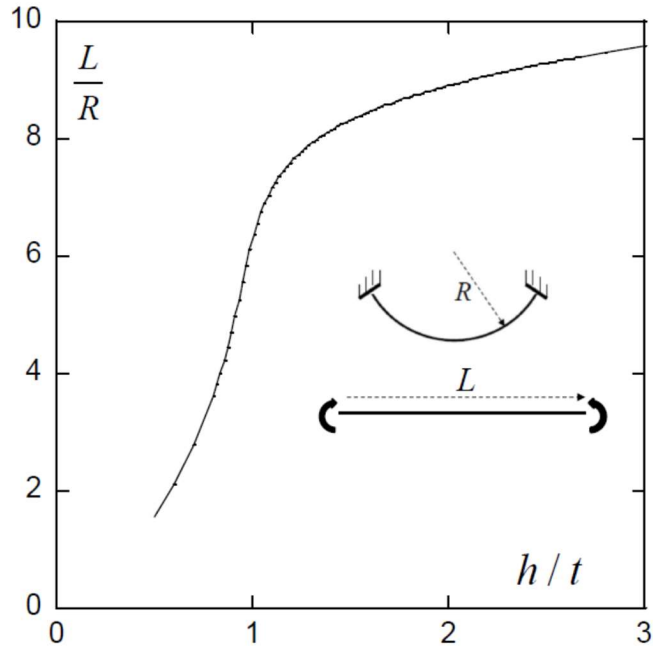

Fig. 5. The stability limit for the length $L$ of a straight rod of rectangular cross-section that can be bent into a clamped circular arc of radius $R$. The height of the cross-section perpendicular to the plane of bending is $h$, the thickness is $t$, and $v = 1/3$ in this plot. For $h/t = 1$, $L/R = 2\pi$.

A clamped bent rod can store relatively large amounts of energy and remain stable. To illustrate this assertion, we compare the elastic energy in a clamped Euler column of length $L$ under axial compression at the critical Euler buckling load (assuming elastic behavior in the compressive state) with the elastic energy stored the clamped straight rod of length $L$ when the natural curvature from which it is formed is at the critical level (3.9). When both entities have the same circular cross-section of radius $r$ and elastic properties, the ratio of these stored energies is

$$\frac{SE_{Euler\,column}}{SE_{straighten\,rod}} = (1+v)\pi^2 \left(\frac{r}{L}\right)^2 \quad (3.15)$$

Slender rods clamped at their ends can store enormously more energy and remain stable when bent than a clamped Euler column in compression. Implications will be discussed later.

*3.3 Michell's Problem: Stability of a circular ring with circular cross-section formed from a twisted straight rod*

In the natural state, the rod is straight. It is subject to a uniform initial twist $d\omega/ds = \omega_0'/R$, with $(\ )' = d(\ )/d\theta$ and $\omega_0'$ as the uniform initial twist, and bent into a one-loop circle with radius $R$, and ends 'welded'. The issue at hand is how much initial twist can be imposed for the circular state to remain stable. This problem was first solved by Michell (1889) and several other solutions were presented in the literature over succeeding years, including a recent elegantly economical solution by Goriely (2006), who also discussed the interesting history of Michell and this problem. Here we add another relatively economical and direct solution based on an investigation of the second variation of the energy. This example illustrates a case in which, due to the initial twist, in-plane and out-of-plane displacements are coupled in the bifurcation mode.

The exact energy functional for the ring from (2.8) and (2.9) is

$$SE/(B_3 R^{-1}) = \tfrac{1}{2}\int_0^{2\pi}\left\{b_2\left(\omega' + (\cos\beta - 1)\gamma' + \sin\beta\cos\gamma\right)^2 + \left(\beta' - \sin\gamma\right)^2 + \left(\sin\beta\,\gamma' - \cos\beta\cos\gamma\right)^2\right\}d\theta \tag{3.16}$$

where $\omega = \alpha + \gamma$, $B_1 = B_3$ and $b_2 = B_2/B_3$. The associated connections are

$$\cos\beta = \sqrt{1 - \varphi_r^2 - \varphi_3^2},\quad \varepsilon_\theta = R^{-1}(u_\theta' + u_r) = \cos\beta - 1 \tag{3.17}$$

$$\varphi_r = R^{-1}u_3' = \sin\beta\cos\gamma,\quad \varphi_3 = -R^{-1}(u_r' - u_\theta) = -\sin\beta\sin\gamma \tag{3.18}$$

The pre-bifurcation state has $\omega' = \omega_0'$, $\beta = 0$, with $SE_0 = \pi R^{-1}B_3(1 + b_2\omega_0'^2)$ and $\gamma$ undetermined. With $\omega' = \omega_0' + \Delta\omega'$ and neglecting terms of order $\beta^3$ and smaller,

$$\frac{SE - SE_0}{B_3 R^{-1}} = \int_0^{2\pi}\left\{b_2\left(\omega_0'\Delta\omega' + \omega_0'\beta\cos\gamma\right) - \left(\beta\sin\gamma\right)'\right\}d\theta + P_2\left(\omega_0',\Delta\omega',\beta,\gamma\right) \tag{3.19}$$

where

$$P_2 = \frac{1}{2}\int_0^{2\pi} \left\{ b_2\left((\Delta\omega' + \beta\cos\gamma)^2 - \omega_0'\beta^2\gamma'\right) + \beta'^2 + \beta^2\gamma'^2 - \beta^2\cos^2\gamma \right\}d\theta \qquad (3.20)$$

The integral in (3.19) that is linear in $\Delta\omega'$ and $\beta$ vanishes because $\Delta\omega$ and $\beta\sin\gamma$ are continuous and periodic, and because $\beta\cos\gamma = R^{-1}u_3'$, neglecting terms of order $\beta^3$ in (3.18). The second integral, $P_2$, which is quadratic in $\Delta\omega'$ and $\beta$, determines the stability of the circular state. The circular state is stable for a pre-twist, $\omega_0'$, if $P_2 > 0$ for all non-zero admissible functions $(\Delta\omega, \beta, \gamma)$, excluding rigid body motions. The limit of the stable states is determined by the smallest value of $|\omega_0'|$ for which there exists a nonzero admissible $(\Delta\omega, \beta, \gamma)$ such that $P_2 = 0$. Because $\Delta\omega$ appears only in the first term in (3.20), minimizing $P_2$ with respect to $\Delta\omega$ implies that $\Delta\omega' = -\beta\cos\gamma$ such that (3.20) reduces to

$$P_2 = \frac{1}{2}\int_0^{2\pi} \left\{ -b_2\omega_0'\beta^2\gamma' + \beta'^2 + \beta^2\gamma'^2 - \beta^2\cos^2\gamma \right\}d\theta \qquad (3.21)$$

The in-plane and out-of-plane displacements are coupled through $\varphi_3/\varphi_r = -\tan\gamma$, where $\gamma(\theta)$ is unknown.

One approach to this eigenvalue problem would be to work with the ODEs generated by rendering $P_2$ stationary. Here, we will use a direct method by considering periodic sinusoidal displacement fields of the form

$$u_r = Raj\cos j\theta, \quad u_\theta = -Ra\sin j\theta, \quad u_3 = Rb\sin j\theta \quad (j = 2, \infty) \qquad (3.22)$$

The phasing of the in-plane and out-of-plane components is important: the choice $u_3 \propto \cos(j\theta)$ does not lead to bifurcation. These modes satisfy the lowest order inextensionality condition $\varepsilon_\theta = 0$. Modes having $j=0$ and $j=1$ are excluded by inextensionality or because they are rigid-body modes. Neglecting terms of order $\beta^3$ and smaller, one has

$$\beta\cos\gamma = \varphi_r = bj\cos j\theta, \quad \beta\sin\gamma = -\varphi_3 = -a(j^2-1)\sin j\theta \qquad (3.23)$$

It follows that the terms in the integrand of (3.21) are

$$(\beta \cos \gamma)^2 = (bj)^2 (\cos j\theta)^2$$

$$\beta^2 = \left(a(j^2-1)\sin j\theta\right)^2 + \left(bj \cos j\theta\right)^2 = (bj)^2 \left((\cos j\theta)^2 + r^2(\sin j\theta)^2\right)$$

$$\beta'^2 = \beta^{-2}\left[\left((a(j^2-1))^2 - (bj)^2\right) j \sin j\theta \cos j\theta\right]^2 = (bj)^2 \frac{\left((r^2-1) j \sin j\theta \cos j\theta\right)^2}{(\cos j\theta)^2 + r^2 (\sin j\theta)^2} \quad (3.24)$$

$$\beta^2 \gamma' = -ab(j^2-1) j^2 = -(bj)^2 jr$$

$$(\beta\gamma')^2 = \beta^{-2}\left(a(j^2-1)bj^2\right)^2 = (bj)^2 \frac{j^4 r^2}{(\cos j\theta)^2 + r^2 (\sin j\theta)^2}$$

where $r = a(j^2-1)/(bj)$. Integrals of the terms in (3.21) are straightforward except for the 2nd for which the following was used

$$(1/\pi)\int_0^{2\pi} (\cos^2 \psi + r^2 \sin^2 \psi)^{-1} \sin^2 \psi \cos^2 \psi \, d\psi = (1+\sqrt{r^2})^{-2}$$

Evaluating $P_2$, one obtains (with the same order of terms in (3.21)):

$$P_2 = \frac{\pi}{2}(bj)^2 \left\{ 2b_2 \omega_0' jr + j^2(1-\sqrt{r^2})^2 + 2j^2 \sqrt{r^2} - 1 \right\} \quad (3.25)$$

which simplifies to

$$P_2 = \frac{\pi}{2}(bj)^2 \left\{ 2b_2 \omega_0' jr + j^2(1+r^2) - 1 \right\} \quad (3.26)$$

The lowest non-trivial eigenvalue is associated with $j = 2$ and given by $b_2 \omega_0' = \sqrt{3}$ with $r = -\sqrt{3}/2$ (or, by $b_2 \omega_0' = -\sqrt{3}$ with $r = \sqrt{3}/2$), in agreement with the original result of Michell (1889) and others as described in the history of the problem by Goriely (2006). Coupling between the in-plane and out-of-plane displacements is essential with $b = \pm\sqrt{3}a$ depending on the sign of $\omega_0'$.

The ring in a 3-loop state is expected to be unstable for any initial twist because, as seen in Section 3.1, such a ring would be unstable without any twist. This can indeed be confirmed by replacing $j\theta$ by $j\theta/N$ in the mode (3.20) and re-evaluating $P_2$. The outcome is $N$ times the righthand side of Eq. (3.26) with $j$ replaced by $j/N$. Then, with $N=3$, it is readily established that $P_2$ can be negative for any value of $\omega_0'$ when $j=1$ or $2$.

## 4. A model problem: The stability of a straight bi-rod formed from two rods with equal natural curvatures

The following problem will illustrate the approach used in the next section for more complicated geometries, but it is of interest on its own merits. The problem is defined in Fig. 6. Apart from the location of their openings, the rods are identical with uniform bending stiffnesses $B_1$ and $B_3$ about the 1- and 3-axes, and torsional stiffness $B_2$ about the 2-axis. Denote the natural curvature of the upper rod by $\kappa_n = \kappa_0$ and that of the lower by $\kappa_n = -\kappa_0$, as defined relative to the $x_2$-axis. The rods are bent until they are straight and then their ends are welded together. The straight configuration of the bi-rod is an equilibrium state with equal and opposite moments in the rods of magnitude, $B_3\kappa_0$, about the 3-axis. The rods are assumed to be inextensional, each having length $L = 2\psi R_0$ where $R_0 = 1/\kappa_0$ and $2\psi$ is the angle subtended by the arc prior to bending. The stability of the straight bi-rod as dependent on $\psi$, or equivalently on $L/R_0$, is analyzed.

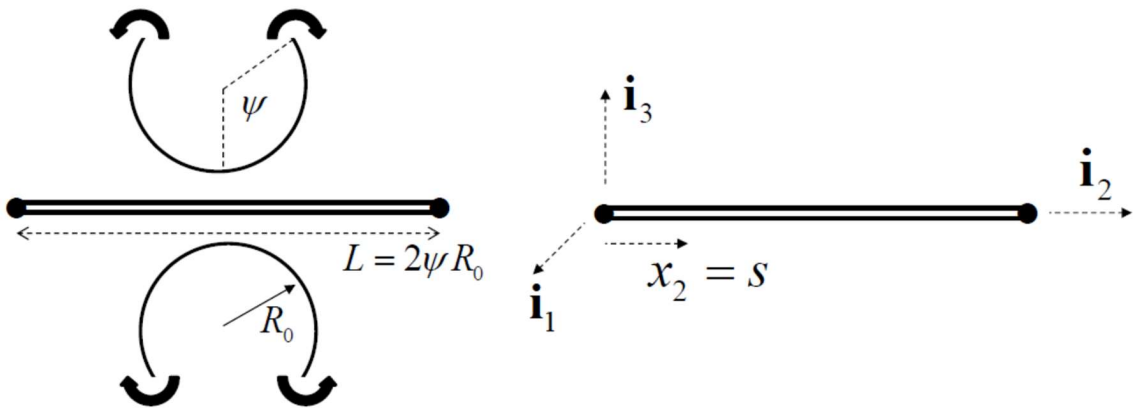

Fig. 6. A bi-rod made of two identical rods of equal and opposite natural curvatures, $\kappa_n = \pm\kappa_0 = \pm R_0^{-1}$, that are bent about the 3-axis into straight segments and welded at the ends.

With the straight configuration as the reference state, the changes of curvature from this state are given by

$$\kappa_1 = \cos\alpha\,\beta' + \sin\alpha\sin\beta\,\gamma', \quad \kappa_2 = \alpha' + \cos\beta\,\gamma', \quad \kappa_3 = \sin\alpha\,\beta' - \cos\alpha\sin\beta\,\gamma' \qquad (4.1)$$

where now $(\ )' = d(\ )/ds$. The linearized strain and rotation variables are

$$\varepsilon_2 = u_2' = \cos\beta - 1, \quad \varphi_1 = u_3' = \sin\beta\cos\gamma, \quad \varphi_3 = -u_1' = -\sin\beta\sin\gamma \qquad (4.2)$$

The strain energy in the bi-rod is

$$SE = \tfrac{1}{2}\int_0^L \{B_1\kappa_1^{(1)2} + B_2\kappa_2^{(1)2} + B_3(\kappa_3^{(1)} - \kappa_0)^2\}ds + \tfrac{1}{2}\int_0^L \{B_1\kappa_1^{(2)2} + B_2\kappa_2^{(2)2} + B_3(\kappa_3^{(2)} + \kappa_0)^2\}ds \qquad (4.3)$$

where the first integral is the energy in the upper rod and the second is that in the lower.

With $\gamma = 0$ and $\alpha = 0$ in the pre-bifurcation state, we evaluate $SE$ in the bifurcated state up to and including terms of order $\alpha^2$ and $\beta^2$. (In-plane displacements, with $\gamma = \pi/2$ and $u_3 = 0$, do not allow bifurcation.) To this same order of accuracy, $\beta = u_3'$, and with this substitution the strain energy is

$$SE = B_3 L \kappa_0^2 + P_2$$
$$P_2 = \tfrac{1}{2}\int_0^L \{B_1 u_3^{(1)''2} + B_2 \alpha^{(1)'2} + 2B_3\kappa_0 \alpha^{(1)'} u_3^{(1)'}\}ds - B_3\kappa_n \alpha^{(1)} u_3^{(1)'}\Big|_0^L \qquad (4.4)$$
$$+ \tfrac{1}{2}\int_0^L \{B_1 u_3^{(2)''2} + B_2 \alpha^{(2)'2} - 2B_3\kappa_0 \alpha^{(2)'} u_3^{(2)'}\}ds + B_3\kappa_n \alpha^{(2)} u_3^{(2)'}\Big|_0^L$$

Because the ends of the rods are welded together, $\alpha$ and $u_3$ must satisfy $\alpha^{(1)} = \alpha^{(2)}$, $u_3^{(1)} = u_3^{(2)}$ and $u_3^{(1)'} = u_3^{(2)'}$ at both ends of the bi-rod. Consequently, the two contributions in (4.4) evaluated at the ends add to zero and can be dropped. The energy in the straight state is $B_3 L \kappa_0^2$, and $P_2$ is quadratic in $\alpha$ and $u_3$. $P_2$ is positive for all admissible non-zero $\alpha$ and $u_3$ (excluding rigid body motions) if the straight state is stable. We seek the upper limit to the stability range, i.e., the minimum value of $\psi$ such that there exists non-zero admissible functions $\alpha$ and $u_3$ for which $P_2 = 0$. The eigenvalue problem for the stability limit requires $\delta P_2 = 0$ for all admissible variations, which, in turn, requires

$$B_1 u_3^{(1)''''} - B_3 \kappa_0 \alpha^{(1)''} = 0, \quad B_2 \alpha^{(1)''} + B_3 \kappa_0 u_3^{(1)''} = 0$$
$$B_1 u_3^{(2)''''} + B_3 \kappa_0 \alpha^{(2)''} = 0, \quad B_2 \alpha^{(2)''} - B_3 \kappa_0 u_3^{(2)''} = 0$$
(4.5)

on $0 \leq x \leq L$, and at the ends,

$$u_3^{(1)''} + u_3^{(2)''} = 0, \quad u_3^{(1)'''} + u_3^{(2)'''} - (B_3 \kappa_0 / B_1)(\alpha^{(1)'} - \alpha^{(2)'}) = 0,$$
$$\alpha^{(1)'} + \alpha^{(2)'} + (B_3 \kappa_0 / B_2)(u_3^{(1)'} - u_3^{(2)'}) = 0$$
(4.6)

in addition to $\alpha^{(1)} = \alpha^{(2)}$, $u_3^{(1)} = u_3^{(2)}$ and $u_3^{(1)'} = u_3^{(2)'}$. Conditions (4.6) are the natural, or dynamic, boundary conditions which ensure there are no external force or moments applied at the ends.

The general solution to (4.5) is

$$u_3^{(1)} = L\left(c_1^{(1)} + c_2^{(1)}(s/L) + c_3^{(1)} \sin(\lambda s / L) + c_4^{(1)} \cos(\lambda s / L)\right)$$
$$\alpha^{(1)} = c_5^{(1)} + c_6^{(1)}(s/L) - K\left(c_3^{(1)} \sin(\lambda s / L) + c_4^{(1)} \cos(\lambda s / L)\right)$$
$$u_3^{(2)} = L\left(c_1^{(2)} + c_2^{(2)}(s/L) + c_3^{(2)} \sin(\lambda s / L) + c_4^{(2)} \cos(\lambda s / L)\right)$$
$$\alpha^{(2)} = c_5^{(2)} + c_6^{(2)}(s/L) + K\left(c_3^{(2)} \sin(\lambda s / L) + c_4^{(2)} \cos(\lambda s / L)\right)$$

with $\lambda = B_3 \kappa_0 L / \sqrt{B_1 B_2}$ and $K = \lambda \sqrt{B_1 / B_2}$. To exclude the possibility of rigid body motions, the left end of the bi-rod is constrained at $x = 0$ such that it cannot undergo displacement or rotation by requiring $\alpha^{(1)} = \alpha^{(2)} = 0$, $u_3^{(1)} = u_3^{(2)} = 0$ and $u_3^{(1)'} = u_3^{(2)'} = 0$. When these conditions, together with the three conditions (4.6), are imposed at $x = 0$, one obtains

$$u_3^{(1)} = L\left(c_3^{(1)}\left(\sin(\lambda s / L) - \lambda s / L\right) + c_4^{(1)}\left(\cos(\lambda s / L) - 1\right)\right)$$
$$\alpha^{(1)} = -K\left(c_3^{(1)}\left(\sin(\lambda s / L) - \lambda s / L\right) + c_4^{(1)}\left(\cos(\lambda s / L) - 1\right)\right) = -(K/L)u_3^{(1)}$$
$$u_3^{(2)} = -L\left(c_3^{(1)}\left(\sin(\lambda s / L) - \lambda s / L\right) + c_4^{(1)}\left(\cos(\lambda s / L) - 1\right)\right) = -u_3^{(1)}$$
$$\alpha^{(2)} = -K\left(c_3^{(1)}\left(\sin(\lambda s / L) - \lambda x / L\right) + c_4^{(1)}\left(\cos(\lambda s / L) - 1\right)\right) = \alpha^{(1)}$$

The final step is to note that by imposing $u_3^{(1)} = u_3^{(2)}$ and $u_3^{(1)'} = u_3^{(2)'}$ at $s = L$, all six conditions at that end will be satisfied. The eigenvalue equations are

$$c_3^{(1)}(\sin\lambda - \lambda) + c_4^{(1)}(\cos\lambda - 1) = 0$$
$$c_3^{(1)}(\cos\lambda - 1) - c_4^{(1)}\sin\lambda = 0$$

requiring $1 - \cos\lambda - (\lambda/2)\sin\lambda = 0$.

The lowest eigenvalue is $\lambda_C = 2\pi$ implying that $c_1^{(4)}$ is the amplitude of the eigenmode with $c_1^{(3)} = 0$ such that

$$u_3^{(1)} = -u_3^{(2)} = Lc_4^{(1)}(\cos(2\pi s/L) - 1), \quad \alpha^{(1)} = \alpha^{(2)} = -K_C c_4^{(1)}(\cos(2\pi s/L) - 1) \qquad (4.7)$$

The critical condition, $B_3 \kappa_0 L / \sqrt{B_1 B_2} = 2\pi$, implies that the largest value of the half-angle of the arc in the natural state such that the bi-rod is stable in the straight state is $\psi_C = \pi\sqrt{B_1 B_2}/B_3$, or equivalently, the longest rods that can form a stable bi-rod is

$$L/R_0 = 2\pi\sqrt{B_1 B_2}/B_3 \qquad (4.8)$$

For rods with a circular cross-section, the limit is $\psi_C = \pi/\sqrt{1+\nu}$. The bifurcation mode, which produces both an opening separation of the bi-rod in the 3-direction and equal rotation in the rods, has no displacement or rotation at the right end. Thus, the solution for each rod is identical for the solution for the stability of a straightened, clamped rod (3.9) in Section 3.2. It is obvious that the solution constructed from (3.10) is an eigenmode. However, it is by no means obvious that this is a mode with the lowest eigenvalue because it is constrained to satisfy $u_3 = 0$ at the right end, while the bi-rod only requires that $u_3^{(1)}(L/2) = u_3^{(2)}(L/2)$. This section shows that it does generate the critical mode because $u_3$ does indeed vanish at the right end of the bi-rod, at least to the order of the bifurcation mode. Furthermore, it is easy to establish that this result also applies if each of the rods in Fig. 6 before bending into the straight configuration is rotated 180 degrees about the horizontal axis with the gap in the upper rod at the bottom and the gap of the lower rod at the top.

Two experimental realizations of steel bi-rods with circular cross-sections are displayed in Fig. 7, one formed from a naturally curved rod with $\psi = \pi/4$, which is well below the stability limit, and the other with $\psi = \pi$, which is about 15% above the stability limit. The ends are clamped by a bracket that also provides some constant separation between the ends of the individual rods. The bi-rod in Fig. 7a is clearly stable in the straight state and requires counter-twisting of the ends to trip it into a stable post-buckled state involving both bending and twisting. Some plastic deformation occurred while twisting the bi-rod and thus the state seen in the righthand photo of Fig. 7a may not be the same if the response had been strictly elastic. The bi-rod in Fig. 7b is unstable in the folded straight state. The figure shows how, starting from a stable state, the bi-rod can be deformed into the straight state, which requires constraint to maintain its stability. When that constraint is removed, the bi-rod springs back to its initial stable state. Photos of the bi-rods in their 'natural state' before both ends are joined are shown in Supplemental Fig. 1 of the Supplementary Materials. Additionally, Video 1 depicts the transition process between states.

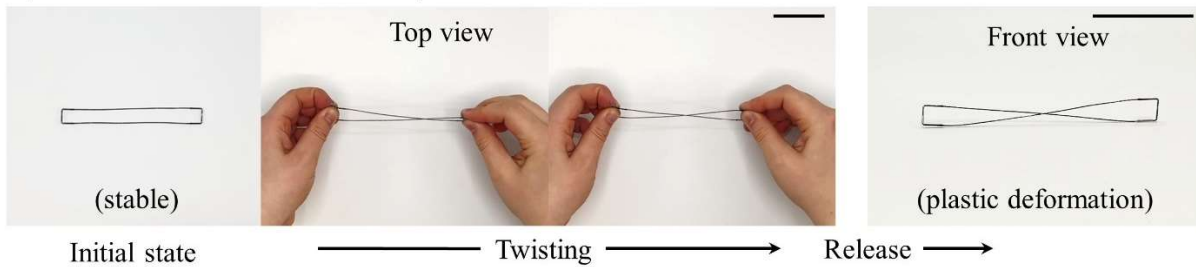

a) Stable straight bi-rod: $\psi = \pi/4$, $R_0 = 75$ mm

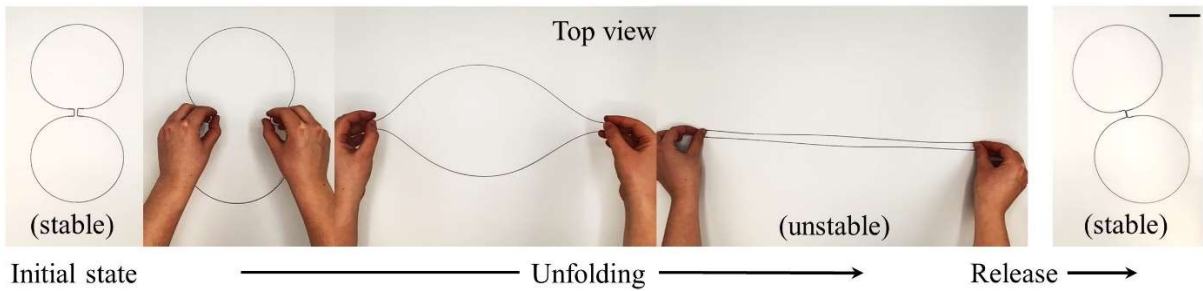

b) Unstable straight bi-rod: $\psi = \pi$, $R_0 = 75$ mm

Fig. 7. Experimental demonstration of bi-rods with circular cross-sections that are stable in a) and unstable in b) in the straight state. a) A stable straight bi-rod with $\psi = \pi/4$ showing the effect of twisting it out of the straight state into a plastically deformed twisted state. b) A bi-rod

with $\psi = \pi$ that is unstable in the straight state. The sequence shows the initial stable non-straight state, manipulation into a constrained straight state, and finally, upon release, return to the initial state. Scale bars: 50 mm.

## 5. Stability of six-sided rings that fold into a straight three-loop

Now consider the ring in Fig. 1a. The ring is formed from six rod segments, each with uniform natural curvature $\kappa_n$, that are bent with a uniform moment, $M_3 = B_3(\kappa_0 - \kappa_n)$, into a 120° arc with radius, $R_0 = 1/\kappa_0$, and 'welded' at the six joints where the arcs form a cusp. This 6-sided 1-loop ring can be twisted into a 3-loop ring with straight sides of length $L = 2\pi R_0 / 3$. The 3-loop configuration is in equilibrium with a uniform bending moment in each segment alternating in sign, $M_3 = \pm B_3 \kappa_n$, from segment to segment. There is no twist in either the 1-loop state or the straight 3-loop state, although twist occurs in the transition from one state to another.[2] We investigate the stability of the straight 3-loop state. The numbering system employed in the stability analysis is shown in Fig. 8a. If the rod segments prior to ring formation are straight with $\kappa_n = 0$, there is no stress in the folded state, and it is clearly stable, but the unfolded one-loop state will have residual stress, and it may or may not be stable—the stability of the unfolded state will not be investigated here. If, on the other hand, the natural curvature of the rod segments is $\kappa_n = 1/R_0$, the one-loop state is unstressed and stable, while the straight folded state is stressed with moments alternating in sign, $M_3 = \pm B_3 \kappa_n = \pm B_3 / R_0$, from segment to segment. We investigate the stability of the straight folded state for rings formed from rods having an arbitrary uniform natural curvature $\kappa_n$.

The elastic energy of the ring with the straight folded state as the reference is

$$SE = \sum_{i=1}^{6} \tfrac{1}{2} \int_0^L \left\{ B_1 \kappa_1^{(i)2} + B_2 \kappa_2^{(i)2} + B_3 \left( \kappa_3^{(i)} - \Delta\kappa_0^{(i)} \right)^2 \right\} ds \qquad (5.1)$$

---

[2] The 3-loop ring is sometimes called a 'covered ring' (Manning and Hoffman, 2006). In forming a 3-loop covered ring, the rod is wound around itself three times and the two ends are lifted slightly to cross-over the ring and joined. There can be a very small twist or non-uniform bending in the vicinity of the cross-over, but this becomes vanishingly small for thinner and thinner rods and is neglected in the rod model.

where $(\kappa_1^{(i)}, \kappa_2^{(i)}, \kappa_2^{(i)})$ are the curvature changes measured from the straight state about the imbedded unit vectors $(\mathbf{e}_1, \mathbf{e}_2, \mathbf{e}_3)$ which coincide with the fixed Cartesian unit vectors $(\mathbf{i}_1, \mathbf{i}_2, \mathbf{i}_3)$ in the straight state. Further, $\Delta\kappa_0^{(i)} = -\kappa_n$ in segments 1, 3 and 5 and $\Delta\kappa_0^{(i)} = \kappa_n$ in segments 2, 4 and 6. The change in sign of $\Delta\kappa_0^{(i)}$ from one segment to the next is due to the flip of the orientation of the segments at each joint in the folded state and our use of the reference coordinate system shown in Fig. 8.

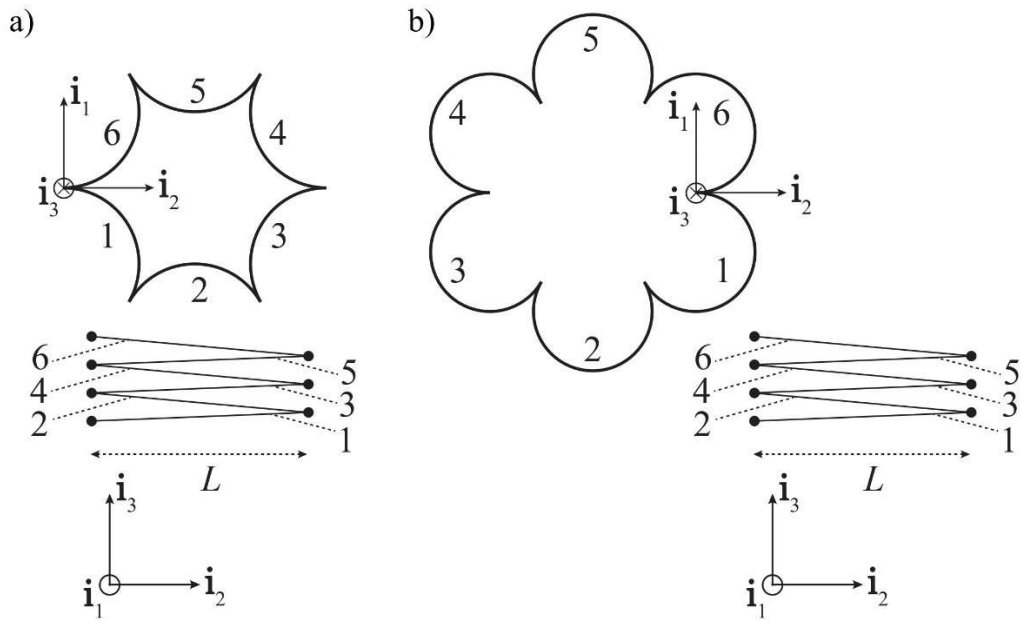

Fig. 8. Numbering system for the segments, coordinate definition, and schematic of the arrangement in the straight folded state. a) Six-sided ring with 120° arc segments. b) Six-sided ring with 240° arc segments. The left ends of segments 1 and 6 are connected in both cases. In the straight folded state, the left ends coincide with $x_2 = 0$ and the right ends with $x_2 = L$.

Following the same procedure described in connection with the bi-rod problem, the strain energy in the $i^{th}$ segment is expressed in terms of $u_3^{(i)}(s)$ and $\alpha^{(i)}(s)$. It is then reduced by retaining all terms up to and including order $u^2$ and $\alpha^2$ such that the strain energy in the $i^{th}$ segment is

$$SE^{(i)} = \tfrac{1}{2} B_3 L \kappa_n^2 + \tfrac{1}{2} \int_0^L \left\{ B_1 u_3^{(i)''2} + B_2 \alpha^{(i)'2} - 2B_3 (\Delta\kappa_0^{(i)} u_3^{(i)'} \alpha^{(i)'}) \right\} ds + B_3 \Delta\kappa_0^{(i)} (u_3^{(i)'} \alpha^{(i)}) \Big|_0^L \quad (5.2)$$

The total strain energy in the ring to this order is $SE = 3B_3 L \kappa_n^2 + P_2$ where the quadratic terms (which sum to the second variation of the system energy about the straight state) are

$$P_2(u_3, \alpha, \kappa_n) = \sum_{i=1}^{6} \tfrac{1}{2} \int_0^L \left\{ B_1 u_3^{(i)''2} + B_2 \alpha^{(i)'2} - 2B_3 (\Delta\kappa_0^{(i)} u_3^{(i)'} \alpha^{(i)'}) \right\} ds \quad (5.3)$$

The geometric conditions at the joints require $(u_3, u_3', \alpha)$ to be continuous across the joints and, because of this, the terms evaluated at the ends of the segments in (5.2) sum to zero. As in the case of the bi-rod problem, the in-plane displacements decouple from $u_3$ and $\alpha$ in the bifurcation problem, and their contribution to the second variation of the energy is positive (apart from rigid body motion) so that bifurcation is controlled by out-of-plane deformation plus twist. If the natural curvature is zero such that $\Delta\kappa_0^{(i)} = 0$, $P_2$ is non-negative and only vanishes for rigid-body displacements. For the present problem, we seek the largest value of $\kappa_n$ such that $P_2 > 0$ for all admissible variations from the straight state, excluding rigid body motions. The critical natural curvature is governed by the eigenvalue problem associate with non-trivial solutions to rendering the first variation of $P_2$ stationary. The differential equations and conditions at the joints that follow from rendering $P_2$ stationary are

$$B_1 u_3^{(i)''''} + B_3 \Delta\kappa_0^{(i)} \alpha^{(i)''} = 0, \quad B_2 \alpha^{(i)''} - B_3 \Delta\kappa_0^{(i)} u_3^{(i)''} = 0, \; i = 1,6 \quad (5.4)$$

$$u_3^{(i)}(L) - u_3^{(i+1)}(L) = 0, \; u_3^{(i)'}(L) - u_3^{(i+1)'}(L) = 0, \; \alpha^{(i)}(L) - \alpha^{(i+1)}(L) = 0$$
$$u_3^{(i)''}(L) + u_3^{(i+1)''}(L) = 0, \; \alpha^{(i)}(L)' + \alpha^{(i+1)}(L)' = 0 \quad (5.5)$$
$$u_3^{(i)'''}(L) + u_3^{(i+1)'''}(L) = -(B_3 \Delta\kappa_0^{(i)} / B_1)(\alpha^{(i)'}(L) - \alpha^{(i+1)'}(L))$$

for $i = 1,3,5$, while for $i = 2,4,6$

$$u_3^{(i)}(0) - u_3^{(i+1)}(0) = 0, \ u_3^{(i)\prime}(0) - u_3^{(i+1)\prime}(0) = 0, \ \alpha^{(i)}(0) - \alpha^{(i+1)}(0) = 0$$

$$u_3^{(i)\prime\prime}(0) + u_3^{(i+1)\prime\prime}(0) = 0, \ \alpha^{(i)}(0)' + \alpha^{(i+1)}(0)' = 0 \tag{5.6}$$

$$u_3^{(i)\prime\prime\prime}(0) + u_3^{(i+1)\prime\prime\prime}(0) = -(B_3 \Delta \kappa_0^{(i)} / B_1)(\alpha^{(i)\prime}(0) - \alpha^{(i+1)\prime}(0))$$

where, when $i = 6$, $i+1$ must be taken as 1. For each set of joint conditions, the first three are geometric requirements and the second three are the natural, or dynamic, conditions that follow from the fact that there are no external forces or moments acting on the joints.

The following dimensionless quantities are employed:

$$\bar{u}^{(i)} = u_3^{(i)} / L, \ \bar{\alpha}^{(i)} = \alpha^{(i)} / (B_3 \kappa_n L / B_2), \ \lambda = B_3 \kappa_n L / \sqrt{B_1 B_2}, \ x = s/L, \ B_2/B_1 \tag{5.7}$$

The eigenvalue is $\lambda$. The details of the following solution to the eigenvalue problem are provided in the Supplementary Materials. The dimensionless equations can be solved analytically for the relation between the unknown quantities at the ends of the six segments. With $\left[u^{(i)}\right] = (\bar{u}^{(i)}, \bar{u}^{(i)\prime}, \bar{\alpha}^{(i)}, \bar{u}^{(i)\prime\prime}, \bar{u}^{(i)\prime\prime\prime}, \bar{\alpha}^{(i)\prime})$, one obtains

$$\left[u^{(i)}(1)\right] = T_A \left[u^{(i)}(0)\right], \ i = 1,3,5 \ \& \ \left[u^{(i)}(1)\right] = T_B \left[u^{(i)}(0)\right], \ i = 2,4,6 \tag{5.8}$$

where $T_A$ and $T_B$ are $6 \times 6$ matrices which depend only on $\lambda$. The conditions at the joints can be expressed as

$$\left[u^{(i+1)}(1)\right] = D_A \left[u^{(i)}(1)\right], \ i = 1,3,5 \ \& \ \left[u^{(i+1)}(0)\right] = D_B \left[u^{(i)}(0)\right], \ i = 2,4,6 \tag{5.9}$$

with $i+1$ taken to be 1 when $i = 6$. $D_A$ and $D_B$ are $6 \times 6$ diagonal matrices except for one off-diagonal element, $D_{56}$, which depends on $B_2 / B_1$. These matrices allow one to compute $\left[u^{(6)}(0)\right]$ in terms of $\left[u^{(1)}(0)\right]$. Then imposing the condition at the joint between segment 1 and 6 in (5.9), one obtains

$$(A - I)\left[u^{(1)}(0)\right] = 0 \tag{5.10}$$

where $A = \left(D_B \left(T_B^{-1} D_A T_A\right)\right)^3$ which depends only on $\lambda$ and $B_2 / B_1$. The form of $A - I$ is

$$A - I = \begin{bmatrix} 0 & A^{(1)} \\ 0 & A^{(2)} \end{bmatrix} \tag{5.11}$$

where $A^{(1)}$ and $A^{(2)}$ are $3\times 3$ matrices. The first three equations in (5.10) require

$$A^{(1)} \left[ \bar{u}^{(1)''}(0), \bar{u}^{(1)'''}(0), \bar{\alpha}^{(1)'}(0) \right] = 0 \tag{5.12}$$

which requires the determinant of $A^{(1)}$ to vanish. By carrying out the numerical calculations prescribed above, the smallest value of $\lambda$ for the determinant of $A^{(1)}$ to vanish is found to be $2\pi$, independent of $B_2/B_1$. The solution to (5.12) with $\lambda = 2\pi$ is any multiple of

$$\left[ \bar{u}^{(1)''}(0), \bar{u}^{(1)'''}(0), \bar{\alpha}^{(1)'}(0) \right] = [1, 0, 0] \tag{5.13}$$

Moreover, this solution satisfies

$$A^{(2)} \left[ \bar{u}^{(1)''}(0), \bar{u}^{(1)'''}(0), \bar{\alpha}^{(1)'}(0) \right] = 0$$

Thus, the critical eigenvalue is $\lambda = 2\pi$ such that the largest value of the natural curvature $\kappa_n$ for which the folded straight 3-loop ring is stable is

$$L\kappa_n = 2\pi\sqrt{B_1 B_2}/B_3 \quad \text{with} \quad L = 2\pi R_0/3 \tag{5.14}$$

This is, again, the same result as for the straightened clamped rod (3.9) in Section 3.2. The other three components determining the eigenfunction, $\left[ \bar{u}^{(1)}(0), \bar{u}^{(1)'}(0), \bar{\alpha}^{(1)}(0) \right]$, are the amplitudes of rigid body modes which can be chosen arbitrarily. For our purposes we will take $\left[ \bar{u}^{(1)}(0), \bar{u}^{(1)'}(0), \bar{\alpha}^{(1)}(0) \right] = 0$. The full eigenmode is plotted in Fig. 9 for the components $\bar{u}(s)$ and $\bar{\alpha}(s)$ as a function of the distance $s$ starting from the left end of segment 1, continuing from the right end back along segment 2 to its left end, and then towards the right end of segment 3, etc. The solution for the critical mode is the piecing together of six of the solutions for the clamped rod of length $L$ such that the moment components at each joint are equal and opposite. Note that the axial moment, or torque, in the rods, $M_2 = B_2\alpha'$, vanishes at the joints, c.f., (3.10).

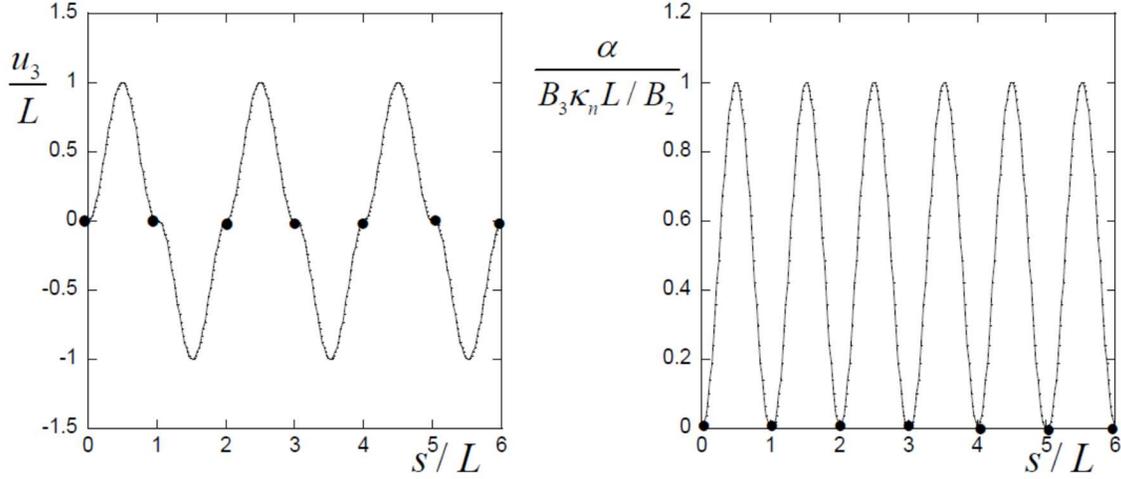

Fig. 9. The bifurcation modal components $u_3$ and $\alpha$ plotted as a function of distance progressing continuously around the ring with six 120° arc segments starting at the left end of segment 1. In each segment $\alpha$ in this plot is the rotation about $\mathbf{i}_2$. The cross-section is circular and $\nu = 1/3$. The joints are tagged with a solid dot. The maximum amplitude of $u_3/L$ is normalized to be unity.

There can be an advantage to constructing this structure from rods with natural curvature (Olson et. al., 2013). If the natural curvature is zero, the elastic energy in the straight folded state is also zero, whereas, with natural curvature, the elastic energy is non-zero and it may be used to help drive the unfolding process. For certain designs, it might even be desirable to have the folded state be unstable. It would be straightforward to restrain the folded state and release the restraint when unfolding is required. The stored elastic energy, which as discussed in Section 3.2 can be substantial, could drive the unfolding process. The maximum natural curvature for stability in (5.14) is proportional to $\sqrt{B_1 B_2}/B_3$ which is plotted for rectangular cross-sections in Fig. 4. Thus, the height to thickness of the cross-section is the important parameter for stability.

Finally, it is noted, without repeating the analysis, that the stability limit for the natural curvature of the folded 6-sided ring formed from 240° arcs in Figs. 1 and 6 is also governed by (5.14). However, the length is twice that for the 120°-arc structure, i.e., $L = 4\pi R_0/3$, reducing the limiting natural curvature for stability by a factor of 2.

An experimental demonstration of the 1- and 3-loop states of the six-sided 120° arc ring is presented in Fig. 10. The cross-section of the rod is rectangular with $h/t = 4$ and the natural curvature of the arcs is $\kappa_n = 1/R_0$ such that the initial unfolded state has no residual stress and is therefore stable (see Supplemental Fig. 2). For this ring, $L\kappa_n = 2\pi/3$, and it is seen from the criterion in (5.14), in conjunction with Fig. 4, that the folded 3-loop state is far below the stability limit, as the experimental realization in Fig. 10 indicates. By applying bending to the ends of the folded ring, it can be brought to an intermediate state at which it snaps dynamically to the unfolded state. The snapping process occurs with vigor consistent with the considerable elastic bending energy stored in the folded state. Photos of the ring before and after the ends of the open ring have been connected are shown in Supplemental Fig. 2 of the Supplementary Materials along with Video 2 showing the unfolding and snapping process.

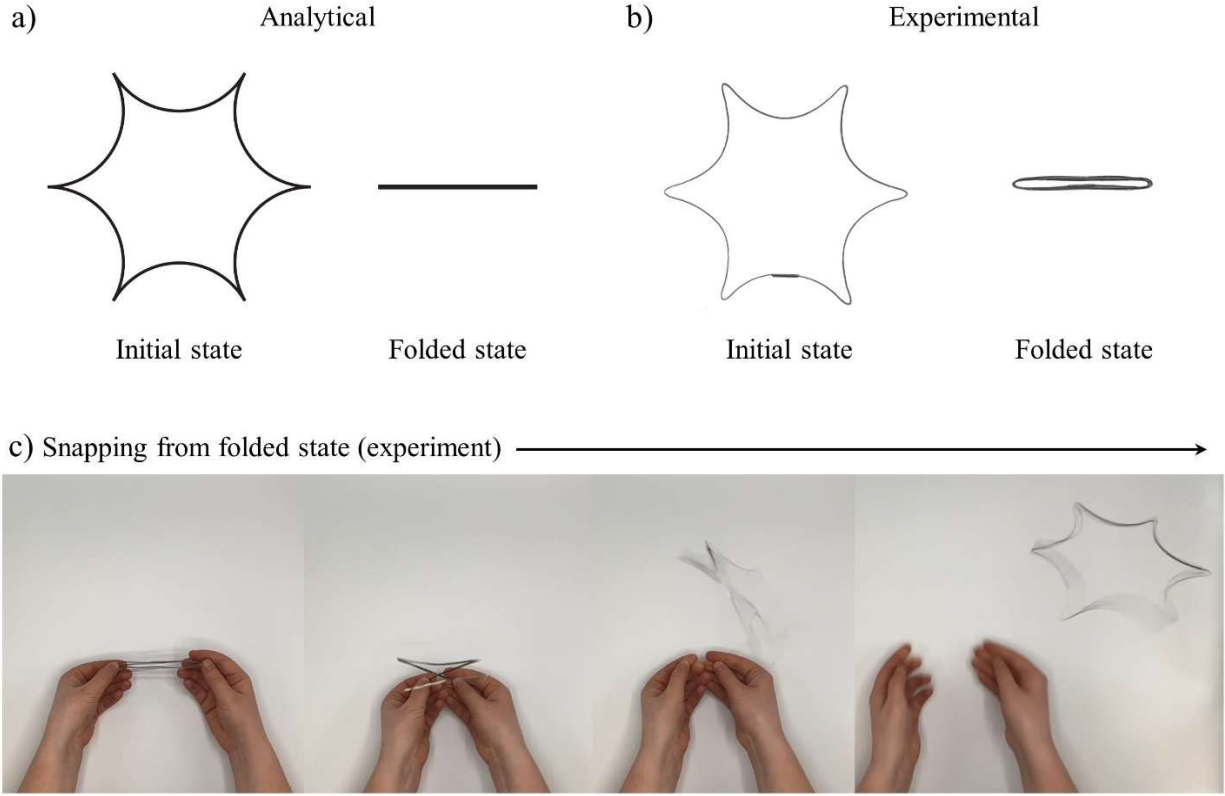

Fig. 10. Unfolded initial (1-loop) and folded (3-loop) states of a six-sided 120° arc ring that is formed from arcs with $\kappa_n = 1/R_0$ such that the 1-loop state has no stored elastic energy. a) The analytical representation. b) The experimental realizations of the two states, both of which are predicted to be stable. c) Applying bending to the ends of the stable folded state until it snaps to the unfolded state.

## 6. Conclusions

The main finding in this paper is the stability limit of the natural curvature for the bi-rod and the straight folded six-sided ring is $L\kappa_n = 2\pi\sqrt{B_1 B_2}/B_3$. The bifurcation mode is a combination of bending of the rods out of the plane of the natural curvature coupled with twist along the length of the rods. The relevance of the fundamental problem in Section 3.2 of the stability limit for bending a clamped rod to reduce its natural curvature has been emphasized. For the case when the rod is bent to become straight, the stability limit on the natural curvature is the same as that for the bi-rod and the straight folded six-sided ring.

The approach to stability used in this paper employs the relations of Kirchhoff rod theory for conservative loads but specialized to a Lagrangian formulation in which the state being investigated is either circular or straight. Stability rests on the positive definiteness of the second variation of the energy of the system. This paper has been concerned only with the ranges of stability and instability and the bifurcation mode from the reference state. Post-bifurcation behavior has not been considered in this paper for any of the problems, although the formulation permits extension into the post-buckling range and is particularly well suited for carrying out studies based on Koiter's (1945) approach to initial post-buckling behavior and imperfection-sensitivity. It should be mentioned that numerical post-bifurcation results for the 1-loop and 3-loop circular ring problems have been presented by Goto et al. (1992), Mouthuy et al. (2012), and by Audoly and Seffen (2015) who deploy the natural curvature as a control parameter to follow quasi-static equilibrium transition paths between the two circular states.

Although the demonstrated experiments on the bi-rod and the six-sided straight folded ring are limited in number and far from being comprehensive, they do suggest that, unlike many structures constructed from rods or beams, these entities may display highly unstable post-bifurcation behavior and strong imperfection-sensitivity. It has been noted that significant elastic energy can be stored in structures such as the bi-rod or straight folded rings constructed by straightening initially curved rods. Future investigations of post-buckling behavior and imperfection-sensitivity of these structures, whether based on the present formulation or the classical Kirchhoff formulation, are likely to be interesting.

**Acknowledgement**

S.L. and R.R.Z. acknowledge funding support from NSF CPS-2201344 and NSF Career Award CMMI-2145601. The authors acknowledge useful discussions with Mr. Jize Dai and Dr. Lu Lu at Stanford University.

**Appendix A. Supplementary materials**

The following is the Supplementary materials to this article.

Video 1: Bi-rods with circular cross-section

Video 2: Six-sided 120º arc ring with rectangular cross-section